\documentclass[12pt]{article}

\usepackage{graphicx}
\usepackage{color}
\usepackage[colorlinks=true, urlcolor=blue, linkcolor=blue, 
citecolor=blue]{hyperref}
\usepackage{bm}
\usepackage{mathtext}
\usepackage{amssymb, amsmath}

\usepackage{epsfig}
\usepackage{bm}
\usepackage[usenames,dvipsnames]{xcolor}
\usepackage{color}

\begin{document}
\thispagestyle{empty}

\begin{center}

\textbf{\Large Diffuse neutrinos from extragalactic supernova remnants: Dominating the 100~TeV IceCube flux}

\vspace{50pt}
Sovan Chakraborty and Ignacio Izaguirre
\vspace{16pt}

\textit{Max-Planck-Institut f\"ur Physik (Werner-Heisenberg-Institut),}\\
\textit{F\"ohringer Ring 6, 80805 M\"unchen, Germany}\\
\vspace{16pt}


\end{center}

\vspace{30pt}


\begin{abstract}
IceCube has measured a diffuse astrophysical flux of TeV--PeV neutrinos. The most
plausible sources are unique high energy cosmic ray accelerators like hypernova remnants (HNRs) and remnants from gamma ray bursts
in star-burst galaxies, which can produce primary cosmic rays with the required energies and abundance. 
In this case, however, ordinary supernova remnants (SNRs), which are far more abundant than HNRs, produce a comparable or larger
neutrino flux in the ranges up to 100--150~TeV energies, implying a spectral break in the IceCube signal around these energies. 
The SNRs contribution in the diffuse flux up to these hundred TeV energies provides a natural baseline and then constrains the expected PeV flux.
\end{abstract}


\newpage
\section{Introduction}
\label{sec:intro}
The IceCube (IC) Collaboration has found 
evidence of a flux of extraterrestrial neutrinos 
in the analysis of the 3-year combined 
data~\cite{Aartsen:2013jdh,Aartsen:2014gkd,Aartsen:2014muf}.
Although, the exact origin of these events is yet to be settled, the signal seems to be incompatible with the 
atmospheric neutrino background and only a small fraction of these events 
is consistent with a galactic origin~\cite{Ahlers:2013xia}. The flux can be reasonably considered to be nearly 
isotropic~\cite{Aartsen:2014gkd} and can be interpreted as the `smoking gun' signal for hadronic processes
~\cite{Gaisser:1994yf,Halzen:2002pg} in the cosmic ray (CR) accelerators. Indeed, such neutrinos would also put 
extremely stringent constraint on neutrino Lorentz-invariance violation~\cite{Borriello:2013ala}. 

Several studies connecting these IC neutrino events
with different cosmic ray production sources estimated the diffuse TeV--PeV background 
(see ~\cite{Anchordoqui:2013dnh} for review and references therein). The general idea being that the CRs loose energy via the hadronic 
processes ($p \gamma$ or $pp$ collisions) with the resultant meson decay to generate neutrinos 
and gamma rays. However, the explanations are based on several assumptions and free parameters,
also due to the poor statistics of events too many explanations seem compatible and is not leading to 
any reasonable conclusion.

Among these possibilities the explanations that these
events are connected to the diffuse background of neutrinos from extreme  high energy accelerators like semi-relativistic 
hypernova remnants (hereafter `HNRs')
~\cite{He:2013cqa,Murase:2013rfa,Liu:2013wia}, long GRB remnant~\cite{Dado:2014mea} embedded in giant molecular cloud 
or slow jet active galactic 
nuclei (AGN)~\cite{Tamborra:2014xia} with dominant contribution from star forming galaxies seem to be highly plausible.
Such objects in star forming galaxies can be tuned to explain the IC diffuse neutrino flux normalization and 
the characteristic sharp cut off at PeV energies. In particular, the $\gamma$ rays generated in the same hadronuclear 
processes would also populate the diffuse gamma ray background and the observed Fermi diffuse background
~\cite{Ackermann:2012vca} can constrain these models~\cite{Liu:2013wia,Tamborra:2014xia}. 

Among the stellar remnants supernova remnants (SNRs) can accelerate protons to PeV energies~\cite{Gaisser}, whereas the HNRs and GRB remnants, 
due to their greater explosion energy and shock velocity, can generate protons up to hundreds of PeV \cite{Wang:2007ya}.
The secondary neutrinos carry almost 5$\%$ of the
parent proton energy, implying that the normal SNRs can contribute to neutrino fluxes 
in the 100~TeV energy range whereas the more energetic CR accelerators may generate neutrinos to PeV energies.
The far away, metal poor galaxies with large star formation rate and large proton densities 
are considered to be strong sources of such remnants \cite{Prieto:2007yb}. In addition, the large pion production efficiency in these 
star burst galaxies (SBGs) makes them excellent sources of the diffuse secondary neutrinos~\cite{Loeb:2006tw}.
The rate of the SBGs increases with the redshift and will have the dominant contribution from high 
redshifts ($z\sim$ 1--2). Therefore one needs to consider the redshift integrated diffuse neutrino 
spectra, which depends on the proton accelerating power of the remnant in the host galaxy and the rate of such remnants. 
The normal SNRs are expected to contribute to this diffuse neutrino flux up to the 100 TeV energy range and the remnants with 
greater ejection energies in the whole TeV--PeV range.

However these extreme high energy remnants like hypernovae and GRBs are rare events, with a local rate of less than 
1$\%$ of the normal SNRs \cite{Guetta:2006gq,Wanderman10}. The SNRs being 100 times more abundant would have larger or at least 
similar energy budget than the HNRs. Thus the normal SNR component of the diffuse neutrinos in the 100~TeV energies should be at 
least of the same order as the HNRs. Also the fact that IC is observing reasonable number of events below the 100 TeV 
energies~\cite{Aartsen:2014muf} emphasizes the importance of the normal SNR contribution to the diffuse flux.

In the following, we calculate the diffuse TeV neutrino flux for SNRs. For the TeV--PeV diffuse flux we take the 
example of the HNRs. 
Our calculations show that for a similar population of host galaxies the neutrino flux up to 100--150 TeV is dominated 
by the {\it{softer}} spectra of the SNRs component and also has a substantial contribution from the HNRs. Whereas, the higher energy 
part (100--1000 TeV) is only HNR generated and has a {\it{harder}} spectra, with a steep cut off around PeV energies. 
Therefore, the diffuse flux would have different spectral behavior in different energy ranges, implying a spectral break around 100~TeV. 
The present statistics of the IC data is insufficient to confirm or exclude this picture and future IC data should test this energy dependent 
spectral behavior. Also, the dominant flux at the 100 TeV energies will give the normalization of the diffuse flux 
and can put interesting constraints on the connection between SNR and HNR rate.

The paper is outlined as follows. In section~\ref{sec:DNBSR} we describe the diffuse
neutrino background from SNR and HNR. We give a brief overview of the uncertainties
which go into the calculation. In the next section~\ref{sec:results} we show the results and
compare the flux with the IC events. Finally in section~\ref{sec:conclusions} we conclude
with a review of our results.
\section{Diffuse neutrino background from stellar remnants}
\label{sec:DNBSR}
The observed diffuse flux of neutrinos from a particular stellar remnant would have contributions
from different redshifts,
\begin{equation}
\frac{dN(E_{\nu}^{ob})}{dE_{\nu}^{ob}} = \frac{c}{4 \pi H_0} \int_{0}^{z_{max}} \frac{dN (E_{\nu})}{dE_{\nu}}
\frac{R_{SR}(z)~dz}{\sqrt{\Omega_M(1+z)^3+\Omega_\lambda}}\,,
\label{eq:DSRnu}
\end{equation}
where the $R_{\text{SR}}(z)$ is the rate of the stellar remnant (SNR or HNR) and $dN (E_{\nu})/dE_{\nu}$ is the source 
neutrino flux. The observed energy ($E_{\nu}^{\text{ob}}$) is connected to the original energy ($E_{\nu} = (1+z)
E_{\nu}^{\text{ob}}$). The Hubble parameter ($H_{0}$) used is 0.69 $\text{km}~\text{s}^{-1}\text{Mpc}^{-1}$. For the standard 
$\lambda$-CDM cosmology, the matter and dark energy density $\Omega_{\text{M}}$ and 
$\Omega_{\lambda}$ are taken to be 0.27 and 0.73, respectively~\cite{Amsler:2008zzb}. 
The source flux $dN (E_{\nu})/dE_{\nu}$ depends on the host galaxy and thus would depend on the relative rate of the 
different types of galaxies. In particular, the relative population of the SBGs ($f_{\text{SBG}}$) is estimated to be 
 $0.1$--$0.2$ \cite{Rodighiero:2011px,Lamastra:2013lfp,Gruppioni:2013jna} and the rest is considered to be normal star forming (NSFG) ones. Thus, 
 total flux from a particular remnant population is the weighted sum of flux from SBGs and NSFGs.

The contribution form each stellar remnant from a particular type of host galaxy is given by ~\cite{Kelner:2006tc},
\begin{equation}
\frac{dN(E_{\nu})}{dE_{\nu}} = \int_{E_{\nu}}^{\infty} \frac{\eta_{\pi}(E_p)}{\kappa} J_{p}(E_p) 
F_{\nu}(\frac{E_{\nu}}{E_p},E_p) 
\frac{dE_p}{E_p}\,,
\label{eq:SRNnu}
\end{equation} where $\eta_{\pi}$ is efficiency of the pion production, 
$\kappa$
is the inelasticity (0.2), $F_{\nu}$ is the secondary neutrino spectrum and $J_{p}$ is the
primary proton spectrum $\sim E_p^{-2} \exp (-E_p/E_p^{max})$. 
The normalization of the proton spectrum is estimated from the total proton energy ($E_p^{\text{T}}$), which is a fraction of 
the ejected energy of the stellar remnant. In particular, for the HNRs $E_p^{\text{T}}$ is in the
range $5\times 10^{51}$--$10^{52}$ erg~\cite{Kulkarni:1998qk}, whereas for SNRs the total proton energy is expected to be
at least one order lower. The maximum energy ($\text{E}_\text{p}^{\text{max}}$) up to which the parent protons can be 
accelerated is governed by the total ejecta energy, magnetic field and the shock radius of the remnant. 
The HNRs with larger ejecta energy and shock radius are expected to have stronger 
neutrino production with broader spectra reaching much higher energies compared to the normal SNRs.
The $\text{E}_\text{p}^{\text{max}}$ for SNRs and HNRs are considered to be in the range 1--10 PeV and $10^2$--$10^3$ PeV, 
respectively~\cite{Gaisser,Wang:2007ya}.

The pion production efficiency ($\eta_{\pi}$) depends on the 
properties of the host galaxy environment~\cite{Liu:2013wia}. Efficient pion production would require that the 
energy loss time ($\text{t}_{\text{loss}}$) due to  $pp$ collisions is smaller than the proton confinement time 
($\text{t}_{\text{conf}}$), thus the  $\eta_\pi$ is estimated as $\eta_\pi=\text{min}(1,\text{t}_{\text{conf}}/\text{t}_{\text{loss}})$.
Whether $\text{t}_{\text{conf}}$ is long enough ($>$ $\text{t}_{\text{loss}}$) to loose energy via 
collisions with the interstellar medium (ISM) gas would depend on the gas density and the magnetic 
field strength of the galactic environment. In particular, $\text{t}_{\text{loss}}=[\kappa\sigma_{\text{pp}}n_{\text{p}}c]^{-1}$, where $\sigma_{\text{pp}}$ is the inelastic nuclear collision
cross section~\cite{Kelner:2006tc}. Clearly, because of this inverse dependence on the ($n_\text{p}$), larger proton dense environment will have smaller energy loss time. 
Therefore, SBGs, with larger proton density ($n_\text{p} \sim$ few $10^2 \text{cm}^{-3}$)~\cite{Tacconi:2005nx} compared to the NSFGs 
($n_\text{p} \sim 10~\text{cm}^{-3}$) ~\cite{Daddi:2009pc,Law:2011de}, are expected to have a more efficient energy loss. 

The $\text{t}_{\text{conf}}$ depends on the scale height of the galaxies (h), the diffusion coefficient(D) and the galactic wind velocities ($\text{V}_{\text{wind}}$). 
At low energies $\text{t}_{\text{conf}}$ is dominated by the advective escape~\cite{Lacki:2009zg} via galactic wind ($\text{t}_{\text{conf}}=\text{t}_{\text{adv}}=\text{h}/\text{V}_{\text{wind}} $) 
and in the higher energies diffusive escape takes the lead ($\text{t}_{\text{conf}}=\text{t}_{\text{diff}}=\text{h}^{2}/4\text{D}$)~\cite{Abramowski:2012xy}. The energy range ($E_{B}$) where both processes 
become competing would result in a break of the spectra and for energies above this break the diffusive escape would give a 
softer spectra with a cutoff-like feature. In particular, the effective spectral shape of the neutrinos in the advection dominating regime broadly follow the primary proton spectra ($E^{-2}$). At higher energies 
the energy dependence  of the diffusion coefficient ($D\propto E^{-0.3}$) would make the resultant neutrino spectra softer. In addition, at high energies, the $\sigma_{pp}$ would also 
have an effect on the overall neutrino spectral shape.

The break energy ($E_{B}$) would also depend on the galactic properties. For 
SBGs the break would appear at much higher energies (PeV) compared to the NSFGs (few hundred GeV) \cite{Murase:2013rfa}
implying the fact that the fluxes in the TeV--PeV energy range would be dominated by SBGs. 
In the advective escape regime the galactic wind velocities in SBGs are much larger than the NSFGs due to the higher rate of supernova explosions, so that one may expect a value of 1500 $\text{km}~\text{s}^{-1}$ 
and 500 $\text{km}~\text{s}^{-1}$ for SBGs and NSFGs, respectively. Regarding the diffusive escape, the high redshift SBGs are expected to have a lower diffusion normalization ($10^{27} \text{cm}^{2} \text{s}^{-1}$) 
because of the higher magnetic fields, whereas for the NSFGs 
we assume a similar diffusion normalization ($10^{28} \text{cm}^{2} \text{s}^{-1}$) to the one of our galaxy~\cite{Trotta11}.  Also the scale height of the NSFGs ($\sim1\text{kpc}$) is larger compared to the SBGs 
($\sim0.5 \text{kpc}$). Adding all these facts, the overall pion production efficiency of the protons is much larger in SBGs, resulting in larger neutrino fluxes  compared to the NSFGs. 
For example, at 100 TeV the $\eta_{\pi}$ is
around $0.02$ and $0.6$ for the NSFGs and SBGs, respectively.

The total high energy diffuse neutrino background would also depend on 
the relative population of the different stellar remnants. 
In particular, the $R_{\text{SR}}(z)$  follow the star formation
history $R_{\text{SFR}}(z)$, for normal SNRs $R_{\text{SNR}}(z)= 1.22 \times 10^{-2} R_{\text{SFR}}(z)
\text{M}_{\odot}^{-1}$~\cite{Baldry:2003xi}. 
The local hypernovae
rate is about 1$\%$ of the supernovae rate~\cite{Guetta:2006gq,Wanderman10,Bhattacharjee:2007gc} and in our analysis we 
consider a HNR rate
$R_{\text{HNR}}(z) \leq 10^{-4} R_{\text{SFR}}(z) \text{M}_{\odot}^{-1}$. For the $R_{\text{SFR}}(z)$ we use the 
concordance models of ~\cite{Hopkins:2006bw,Yuksel:2006qb}, where in low z ($< 1$) the SFR increases as $(1+z)^{3.4}$ to 
remain constant till the redshifts $z \sim$ 4. Beyond  $z=4 $ the SFR decreases in extreme 
rapid fashion $\sim (1+z)^{-7}$ and for the local star formation rate $R_{\text{SFR}}(0)$, we use $10^{-2} 
\text{M}_{\odot} \text{yr}^{-1} \text{Mpc}^{-3}$.

\section{Results}
\label{sec:results}
The flux of high energy diffuse neutrino background depends on the considerations
described in the previous section. To study the dependence on the uncertainties of the
input parameters we calculate the flux for the extreme ranges of the parameters. In
particular, the input parameters like total proton energy ($E_p^{\text{T}}$) and the relative population 
($f_{\text{SBG}}$) of the SBGs in comparison to the NSFGs 
are difficult to estimate accurately. For the HNRs we use $E_p^{\text{T}}$ and $f_{\text{SBG}}$ in the range
$5\times 10^{51}$--$10^{52} $ erg and 0.1--0.2, respectively. In case of the SNRs the
$E_p^{\text{T}}$ is considered in the $5\times 10^{50}$--$10^{51} $ erg range. Also, there will be
uncertainties in the local HNR rates. The value used in the analysis is simply the upper limit.
However, in the SBGs the HNRs may have a higher rate compared to the normal star forming galaxies \cite{Yuksel:2006qb,Yuksel:2008cu}. 
Regarding the $\text{E}_\text{p}^{\text{max}}$ we take 5 PeV and 1 EeV for the SNRs and HNRs, respectively.
\begin{figure}[!t]
\begin{center}
 \includegraphics[angle=0,width=0.6\textwidth]{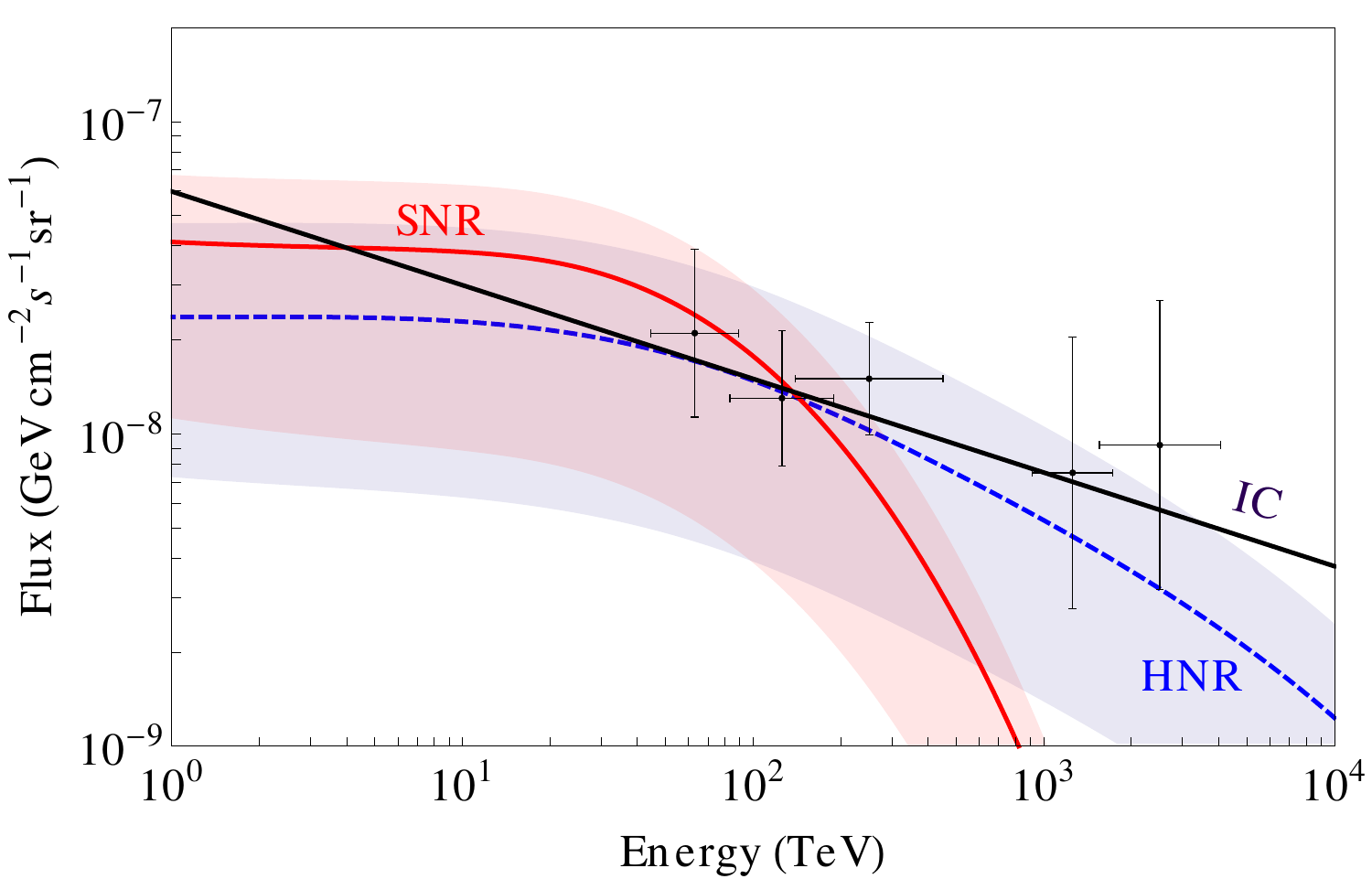}
\end{center}
\caption{ Single flavor diffuse neutrino spectra from both the SNRs (red
continuous curve) and HNRs (blue dashed) contributions. The estimates have contribution
form both the normal and the star burst galaxies. The band describes the
uncertainties in the input parameters. The IC best fit spectra and the data is given by the black
continuous curve and black points, respectively~\cite{Aartsen:2014gkd}.}
\label{fig-componentflux}
\end{figure}

Fig.~\ref{fig-componentflux} shows the single flavor diffuse neutrino spectra ($E_{\nu}^2dN/dE_{\nu}$) for both the SNRs (red
continuous curve) and HNRs (blue dashed) contributions. For both the SNR and HNR curve $f_{\text{SBG}}$ is taken as 0.15. The $E_p^{\text{T}}$ is $5\times 10^{50}$ erg and $5\times 10^{51}$ erg for the SNR and HNR, 
respectively. The bands around the curves describe the uncertainties in $f_{\text{SBG}}$ and $E_p^{\text{T}}$.
The IC data with error bar is given by black points and the best fit spectra 
($E_{\nu}^2dN/dE_{\nu}=1.5\times10^{-8}(\text{E}/100\text{TeV})^{-0.3}~\text{GeV}\text{cm}^{-2}\text{s}^{-1}\text{sr}^{-1}$)
~\cite{Aartsen:2014gkd} is given by the black thick curve.
Both the HNR and SNR estimates have contributions from the normal and the star burst galaxies. 
Due to high $\eta_\pi$ and $E_B$ in the TeV--PeV energies the SBGs contribution dominates over the normal ones. 
The HNR contribution steeply decreases around few hundred TeV to give a sharp cut off around PeV 
energies. This feature of the diffuse HNR neutrino flux makes it an excellent candidate to the observed IC  
events extended to PeV energies. However, in the lower TeV energies the same galaxy population is also giving 
a strong diffuse flux coming from the normal SNRs. The SNRs contribution is dominating up to hundred TeV 
energies. The limitation of the normal SNRs to accelerate protons above PeV energies translates into this 
cutoff at around 100 TeV in the secondary neutrino spectra. 

\begin{figure}[!t]
\begin{center}
 \includegraphics[angle=0,width=0.6\textwidth]{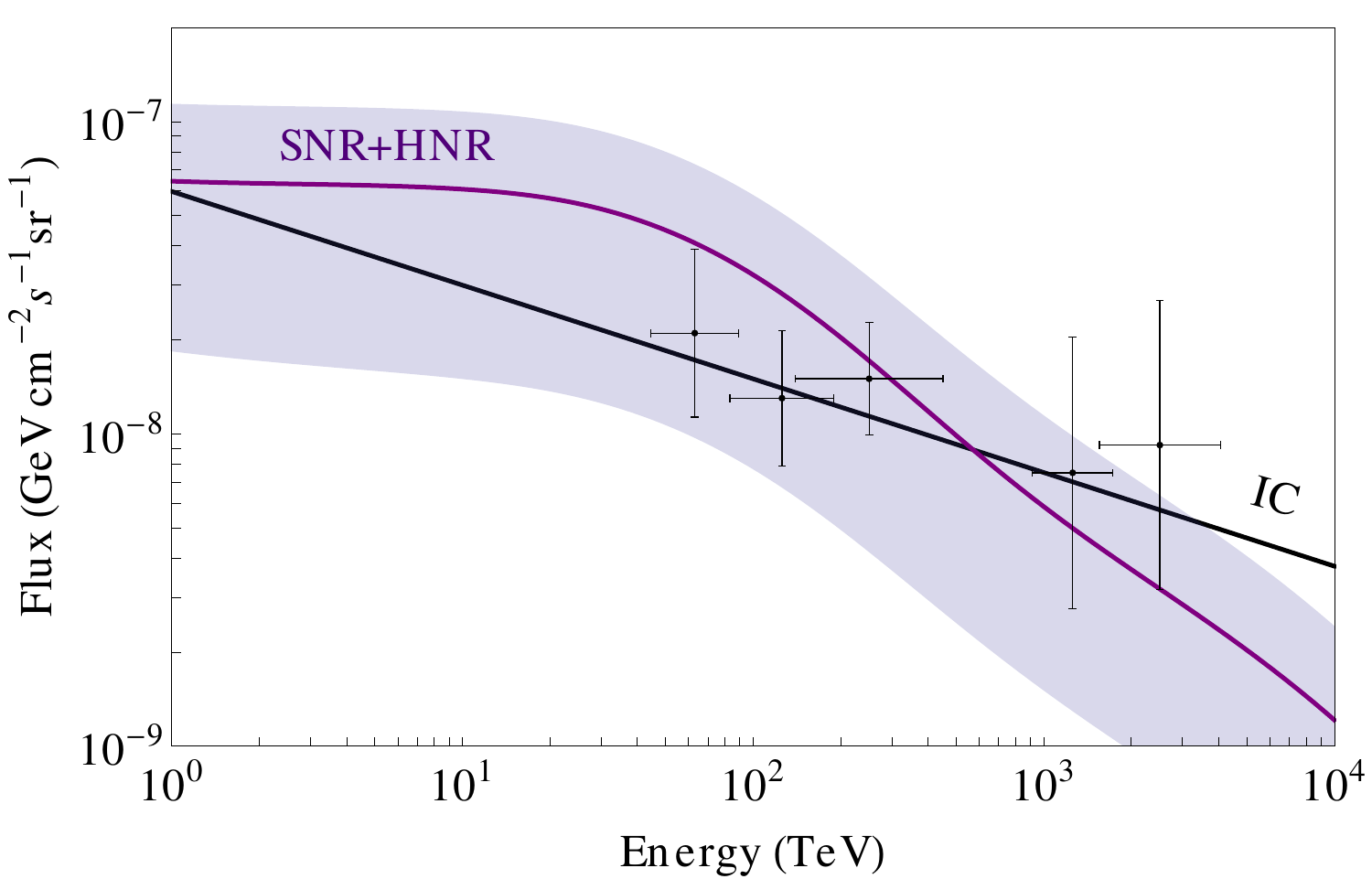}
\end{center}
\caption{ The total diffuse neutrino flux from both stellar remnants is given by the purple curve. The band 
describes the uncertainties in the input parameters. The IC data is the same as in fig~\ref{fig-componentflux}.}
\label{fig-totalflux}
\end{figure}

The total diffuse neutrino flux from both types of stellar remnants and all galaxy population is given in the 
Fig.~\ref{fig-totalflux}. The purple curve is the total diffuse flux corresponding to the SNR and HNR curves in 
Fig.~\ref{fig-componentflux}. The band around the total flux curve describes the uncertainties in the input 
parameters. The IC best fit and the data are again plotted by the black curve and the points, respectively. 
Up to the 100 TeV energies the spectral shape of the diffuse neutrino flux is dominated by the SNRs contribution.
As the SNRs contributions rapidly decrease around 100 TeV, the higher energy contribution 
is solely from the HNRs. The HNRs flux in the higher energies is harder compared to the softer SNR 
contribution. Thus the total diffuse neutrino spectra will have a break in around 100 
TeV energies. This structure in the spectra would be a generic feature of 
these models and can put interesting constraints. Indeed, the uncertainties in the 
model parameters can only effect the overall normalization of the diffuse flux, but the spectral structure remains similar  . 

In particular, the extreme high energy accelerators like HNRs have many uncertainties.
The shock acceleration, the magnetic fields and efficiency of kinetic energy to 
CR energy transfer in these rare objects still need detailed understanding. The relative rate of HNR to
SNR is another ingredient in these uncertainties. However, the SNRs are much 
better understood and their rate is also well estimated. In fact, the MeV neutrinos emitted during the supernova explosion phase of these objects would 
contribute to the diffuse SN neutrino background (DSNB). Together with the peculiar neutrino oscillations in SN, DSNB is one of the very interesting signals 
to study~\cite{Chakraborty:2008zp,Chakraboty:2010sz}. The Super Kamiokande (SK) 
upper limit of DSNB puts a strong limit on the SN rate~\cite{Malek:2002ns}. Thus compared to the uncertain
HNR generated diffuse flux the SNR contribution can be considered as `guaranteed' component of the 
diffuse high energy neutrinos. For any high energy neutrino observation this dominant 
and guaranteed SNR contribution would be the natural baseline to find the diffuse flux 
normalization in the sub 100 TeV energies. In fact, as pointed out in Fig.~\ref{fig-totalflux}, extending these normalization to 
higher energies will put strong limits on the HNR contribution and they may become weaker to explain the IC events 
in the PeV energies. The observed IC data shows a mild preference in the low energies. The present IC 
flux would allow a large fraction of the input parameters in the PeV energies as 
the flux quickly decreases due to diffusion dominated CR acceleration. 
In the low energies up to few 100~TeV the diffuse neutrino parameters are constrained by the present IC observed flux. 
However, the present statistically limited IC observation can not confirm or discard these models. 
Future IC data should reveal if any such spectral feature exists in the isotropic neutrino flux. With 
the present data one can only point out the strong connection between the fluxes of the TeV--PeV 
diffuse neutrino background in different energy ranges.

\section{Conclusions}
\label{sec:conclusions}
Very high energy cosmic rays accelerated by the stellar remnants of objects 
like semi relativistic hypernovae and relativistic GRB remnants interacting with giant 
molecular clouds through hadronic processes are expected to produce diffuse TeV--PeV 
neutrino flux within the reach of IC sensitivities. We found that indeed, the detected 
IC neutrino flux can contain such an explanation with diffuse neutrino flux originating 
in stellar remnants in high redshift, metal poor star burst galaxies. However, the normal 
SNR population in such galaxies also contribute to this diffuse neutrino flux. Our 
estimation of such a diffuse neutrino flux shows that up to 100--150~TeV energies, 
the diffuse background is dominated by the contribution from normal SNRs. 
In higher energies the diffuse flux is dominated by the HNR component.  

Therefore, such a HNR model of diffuse high energy neutrinos successfully explaining the IC
PeV events would have strong constrain from the SNRs contribution form the TeV energies, so that it doesn't overpopulate
the flux below 100 TeV. The present IC data indicates mild tension with a large part of the 
parameter space of these diffuse models. Future IC data should give a conclusive indication from the 
spectral shape of the flux. In fact the results would also depend on the connection between the 
star formation history of the different remnants. In the high redshift galaxies the very high energy 
accelerators are expected to be more abundant compared to the local galaxy population, the absence of 
such a spectral shape would constrain
the relative rate and redshift dependence of the different remnants. Such a detailed study would 
give more exhaustive answers and should be pursued in future.

Furthermore, the same hadronic interactions of the stellar remnant CRs would also generate a diffuse 
gamma ray component from the neutral meson decay and would populate the observed diffuse gamma ray background 
in the 10-100 GeV energies. The corresponding diffuse gamma ray background of the detected IC neutrino 
flux reasonably agrees with the Fermi limits \cite{Murase:2013rfa,Tamborra:2014xia}. Thus, such a model of stellar 
remnants producing the diffuse neutrino spectra in star forming galaxies would also get constrained 
from the Fermi limits. The ordinary SNRs are more susceptible to overpopulate the Fermi limits \cite{Liu:2013wia} 
and such a detailed study with different model parameters would have important multi-messenger implications. 
However, such a limit would also involve more parameters as one would need to understand the cascading 
of high energy gamma rays to the low energy flux and hence more uncertainties in the already parameter abundant model.
The constraints coming only from the future IC neutrino data would be 
independent of uncertainties connecting the diffuse neutrino background with the diffuse gamma ray 
flux and should give the cleanest limits.     

The present analysis shows the importance of the multi energy study involving the SNRs. We focus 
on the HNRs as the higher energy counterpart. However, there are several other possibilities
from relativistic GRB remnants to slow jet AGNs. The relative population and SFR of these 
rare, high energy cosmic accelerators would need more understanding. On the other hand the nature 
and the rate of the normal SNRs are better understood and the diffuse neutrino flux from SNRs 
is more natural compared to the extreme high energy accelerators contribution.
Hence the SNRs contribution up to hundreds of TeV energies should give the reference and baseline for the 
fluxes of all the {\it{connected}} higher energetic astrophysical events.\\	
\textit{Note added}: A week after the submission of the present work,~\cite{Senno:2015tra} 
proposed a similar idea to the one presented here.

\section*{Acknowledgments}
We thank Georg Raffelt and Irene Tamborra for useful discussions and remarks.
S.C.\ acknowledges support
from the European Union through a Marie Curie Fellowship, Grant No.\
PIIF-GA-2011-299861 and through the ITN “Invisibles”, Grant No.\ PITN-GA-2011-289442.


\end{document}